\documentclass[aps,preprint,amssymb,amsmath]{revtex4}
\usepackage{amsmath,amssymb}
\usepackage{graphicx}
\usepackage{enumerate}
\newcommand{\AddrLNF}{
  {\it INFN, Laboratori Nazionali di Frascati,C.P. 13, I00044 
    Frascati, Italy}}
\newcommand{\AddrUdeA}{
  {\it Instituto de F\'\i sica, Universidad de Antioquia,
    A.A.{\it{1226}}, Medell\'\i n, Colombia}}
\newcommand{\AddrEIA}{
  {\it Escuela de Ingenier\'{\i}a de Antioquia,
     A.A.{\it{7516}}, Medell\'\i n, Colombia}}
\begin{document}
\preprint{arXiv:0907.0682} 
\title{Decaying neutralino dark matter in anomalous $U(1)_H$ models} 
\author{D. \surname{Aristizabal Sierra}}
\affiliation{\AddrLNF}
\author{D. Restrepo}
\affiliation{\AddrUdeA}
\author{Oscar Zapata}
\affiliation{\AddrEIA}
\begin{abstract}
   In supersymmetric models extended with an anomalous $U(1)_H$
   different R-parity violating couplings can yield an unstable
   neutralino.  We show that in this context astrophysical and
   cosmological constraints on neutralino decaying dark matter forbid
   bilinear R-parity breaking neutralino decays and lead to a class of
   purely trilinear R-parity violating scenarios in which the
   neutralino is stable on cosmological scales. We have found that
   among the resulting models some of them become suitable to explain
   the observed anomalies in cosmic-ray electron/positron fluxes.
\end{abstract}
\maketitle

\section{Introduction}
\label{sec:int}
Recent measurements of high energy cosmic rays reported by different
collaborations have attracted a great deal of attention as, in
contrast to what is expected from spallation of primary cosmic rays on
the interstellar medium, the electron/positron flux exhibits
intriguing features.  The PAMELA collaboration saw a rise in the ratio
of positron to electron-plus-positron fluxes at energies 10-100 GeV
\cite{pamela}.  The ATIC experiment reported the discovery of a peak
in the total electron-plus-positron flux at energies 600-700 GeV
\cite{atic} and more recently the Fermi~LAT \cite{fermi} collaboration reported an
excess on the total electron-plus-positron flux in the same energy
range as ATIC but less pronounced \cite{Bergstrom:2009fa}. These findings in
addition to those published by the HESS \cite{hess}, HEAT \cite{heat},
and PPB-BETS \cite{ppb-bets} experiments might be indicating the
presence of a nearby source of electrons and positrons. Possible
sources can have an astrophysical origin e.g.  nearby pulsars
\cite{Profumo:2008ms} but more interesting they can be related with
either dark matter (DM) annihilation \cite{ADM,Cheung:2009si} or decay
\cite{Cheung:2009si,DDM,Nardi:2008ix,Yin,Gogoladze:2009kv,Chen:2009mj,Shirai:2009fq,Fukuoka:2009cu}.  In particular, decaying DM
scenarios are quite appealing as, in contrast to models relying on DM
annihilation, they are readily reconcilable with the observed
electron-positron excess \cite{Nardi:2008ix}.

In R-parity breaking models the LSP is unstable and its lifetime is
determined by supersymmetric parameters and R-parity breaking
couplings.  Depending on their values, the phenomenological
implications of a decaying LSP can range from collider physics up to
cosmology and astrophysics. In the later case the possibility of a
long--lived, but not absolutely stable LSP, leads to decaying DM
scenarios as was shown in
Refs.~\cite{Barbieri:1988mw,Berezinsky:1996pb,Baltz:1997ar,Gupta:2004dz,Huber:2005iz}
and indeed they have been recently reconsidered as a pathway to
explain the observed anomalies in cosmic--ray electron/positron fluxes
\cite{Yin,Gogoladze:2009kv,Chen:2009mj,Shirai:2009fq}.  So far, most of
the analyses have been carried out by {\it ad hoc} selections of
particular sets of R-parity violating couplings and/or by assuming
tiny couplings. Thus, it will be desirable to build a general
framework for supersymmetric decaying DM in which the allowed
couplings and their relative sizes arise from generic considerations
rather than from {\it ad hoc} choices as was done in
\cite{Shirai:2009fq,Fukuoka:2009cu,Berezinsky:1996pb}. This is the
purpose of this work.

In supersymmetric models extended to include an anomalous horizontal
$U(1)_H$ symmetry {\it \`a la} Froggatt-Nielsen (FN) \cite{Froggatt},
the standard model particles and their superpartners do not carry a
R-parity quantum number and instead carry a horizontal charge
($H$--charge). For a review see \cite{Dreiner:2003hw}.  In addition, these
kinds of models involve new heavy FN fields and, in the simplest
realizations, an electroweak singlet superfield $\Phi$ of $H$--charge $-1$.
R-parity conserving as well as R-parity violating $SU(3)\times
SU(2)\times U(1)_Y\times U(1)_H$ invariant effective terms arise once
below the FN fields scale ($M$) the heavy degrees of freedom are
integrated out.  These terms involve factors of the type $(\Phi/M)^n$,
where $n$ is fixed by the horizontal charges of the fields involved
and determines whether a particular term can or cannot be present in the
superpotential. The holomorphy of the superpotential forbids all the
terms for which $n<0$ and although they will be generated after
$U(1)_H$ symmetry breaking (triggered by the vacuum expectation value of the scalar
component of $\Phi$, $\langle\phi\rangle$) via the K\"ahler potential
\cite{Giudice} these terms are in general much more suppressed than
those for which $n\ge0$.  Terms with fractional $n$ are also forbidden
and in contrast to those with $n<0$ there is no mechanism through
which they can be generated.  Finally, once $U(1)_H$ is broken the
terms with positive $n$ yield Yukawa couplings determined---up to order
one factors---by $\theta^n=(\langle\phi\rangle/M)^n$. The standard
model fermion Yukawa couplings typically arise from terms of this
kind. Correspondingly, supersymmetric models based on an $U(1)_H$
Abelian factor are completely specified in terms of the $H$--charges.

In the case of supersymmetric models based on an anomalous $U(1)_H$
flavor symmetry the quark masses, the quark mixing angles, the charged
lepton masses, and the conditions of anomaly cancellation constrain
the possible $H$--charge assignments
\cite{Leurer:1992wg,Binetruy:1996xk}. Since the number of constraints
is always smaller than the number of $H$--charges some of them are
necessarily unconstrained and apart from theoretical upper bounds on
their values \cite{Casas:1987us} they can be regarded as
free parameters that should be determined by additional
phenomenological input. For this purpose neutrino experimental data
has been used resulting in models in which neutrino masses are explained \cite{Dreiner:2003hw,Choi:1998wc,Mira,Dreiner:2003yr,Dreiner:2006xw,Dreiner:2007vp}.  Here we
adopt another approach by assuming a decaying neutralino as a dark
matter candidate. We will show that astrophysical and cosmological
observations exclude the possibility of having neutralino decays
induced by bilinear R-parity violating couplings and that this in turn
lead to a variety of purely trilinear R-parity breaking scenarios among
which we found models that feature a single trilinear R-parity
breaking coupling (minimal trilinear R-parity violating models) and
that turn out to be suitable to explain the reported anomalies in
cosmic-ray electron/positron fluxes in either models with TeV-ish
supersymmetric mass spectra or split supersymmetry.
 
The rest of this paper is organized as follows: In
Sec.~\ref{sec:charges} we will describe the possible models that arise
as a consequence of the constraints imposed by astrophysical and
cosmological observations on a decaying neutralino as dark matter.  We
will focus on the resulting minimal trilinear R-parity violating
models and discuss some realizations coming from specific $H$--charge
assignments. In Sec.~\ref{sec:pamela-fermi-atic} we will show that
current data on cosmic-ray electron/positron fluxes are well described
by these type of models. Finally in sec. \ref{sec:summary} we will
summarize and present our conclusions.
\section{Minimal R-parity violating model}
\label{sec:charges}
The most general supersymmetric version of the standard model has a 
renormalizable superpotential given by
\begin{equation}
  \label{eq:5}
  W = \mu_\alpha\widehat{L}_\alpha\widehat{H}_u + 
  h^u_{ij}\widehat{H}_u\widehat{Q}_i\widehat{u}_j + 
  \lambda_{\alpha\beta k}\widehat{L}_\alpha\widehat{L}_\beta\widehat{l}_k +
  \lambda'_{\alpha j k}\widehat{L}_\alpha\widehat{Q}_j\widehat{d}_k + 
  \lambda''_{ijk}\widehat{u}_i\widehat{d}_j\widehat{d}_k
  \,,
\end{equation}
where Latin indices $i, j, k, \dots$ run over the fermion generations
whereas Greek indices $\alpha, \beta, \dots$ run from 0 up to 3. In
the notation we are using $\widehat L_0=\widehat H_d$ and the fermion
Yukawa couplings are given by $h^l_{ij}=\lambda_{0ij}$ and
$h^d_{ij}=\lambda'_{0ij}$. Bilinear couplings $\mu_i$ as well as the
trilinear parameters $\lambda_{ijk}$ and $\lambda_{ijk}'$ break lepton
number whereas baryon number is broken by the couplings
$\lambda_{ijk}''$.  When extending a supersymmetric model with a
$U(1)_H$ Abelian factor, the size of all the parameters entering in the
superpotential arises as a consequence of $U(1)_H$ breaking. In
particular, the lepton and baryon number couplings are well suppressed
or can even be absent without the need of
$R$--parity \cite{Dreiner:2003hw,Choi:1998wc,Mira,Dreiner:2003yr,Dreiner:2006xw,Dreiner:2007vp,Choi,Joshipura:2000sn}

These kinds of frameworks are string inspired in the sense that the
anomalous $U(1)_H$ symmetry may be a remnant of a string model
\cite{Choi,Dreiner:2003hw} implying that the {\it natural} scale of
the FN fields $M$ can be identified with $M_P$ and that anomaly
cancellations can proceed through the Green-Schwarz mechanism
\cite{Green}. Below the string scale, the terms in the superpotential
[Eq.~(\ref{eq:5})] as well as the K\"ahler potential are generated
after $U(1)_H$ breaking induced by $\langle\phi\rangle$ and as we
already discussed may be vanishing or suppressed depending on the
$H$--charge assignments of the different fields involved which in string
models are always constrained to be not too large.  Accordingly, in
what follows we will constrain the $H$--charges to satisfy the condition
$|H(f_i)|<10$ that as highlighted in Refs. \cite{Dreiner:2003hw,Choi} leads to a complete
consistent supersymmetric flavor model.

Before proceeding we will fix our notation: Following Ref.
\cite{Mira} we will denote a field and its $H$--charge with the same symbol,
i.e.  $H(f_i)=f_i$, $H$--charge differences as $H(f_i-f_j)=f_{ij}$ \cite{Dudas:1995yu},
bilinear $H$--charges as $n_\alpha = L_\alpha + H_u$, and trilinear
$H$--charges according to $n_{\lambda_{ijk}}$ with the index determined
by the corresponding trilinear coupling, that is to say the index can
be given by $\lambda_{ijk}$, $\lambda_{ijk}'$, or $\lambda_{ijk}''$.
We fix $\theta=\langle\phi\rangle/M\simeq 0.22$ \cite{Dreiner:2003yr,Irges:1998ax} and $H(\phi)=-1$
without loss of generality. Furthermore we parametrize
$\tan\beta=\theta^{x-3}$ ($x=H_d + Q_3 + d_3 = H_d + L_3 + l_3$) such
that it ranges from 90 to 1 for $x$ running from 0 to 3 (see Ref.
\cite{Mira} for more details).

As already stressed any coupling in the superpotential is determined
up to order 1 factors by its $H$--charge. Thus, any bilinear or
trilinear couplings $\mu_\alpha$ and $\lambda_T$ must be given by
\cite{Dreiner:2003hw,Binetruy:1996xk}
\begin{align}
  \label{eq:2}
  \mu_\alpha\sim &
  \begin{cases}
    M_P\theta^{n_\alpha}         & n_\alpha\ge 0\\
    m_{3/2}\theta^{|n_\alpha|} & n_\alpha<0\\
    0                        & n_\alpha\ \text{fractional}
  \end{cases}
  &\lambda_T\sim&
  \begin{cases}
    \theta^{n_\lambda}               & n_\lambda\ge 0\\
    (m_{3/2}/M_P)\theta^{|n_\lambda|} & n_\lambda<0\\
    0                               &n_\lambda\ \text{fractional}
  \end{cases}\,.
\end{align}
The individual $H$--charges in turn are determined by a set of
phenomenological and theoretical conditions which can be listed as
follows:
\begin{itemize}
\item Eight phenomenological constraints arising from six quark and lepton
  mass ratios plus two quark mixing angles:
  \begin{align} 
    \label{hierarchy}
    m_u:m_c:m_t &\simeq \theta^{\,8}:\theta^{\,4}:1\,, \nonumber \\
    m_d:m_s:m_b &\simeq
    \theta^{\,4}:\theta^{\,2}:1\,,  \nonumber \\
    m_e:m_\mu:m_\tau &\simeq
    \theta^{\,5}:\theta^{\,2} :1\,,   \nonumber \\
    V_{us}\simeq \theta\,, & \quad  V_{cb}\simeq \theta^{\,2}\,.
  \end{align}
  Once imposed they give rise to the constraints (see Ref. \cite{Mira} and
  references therein)
  \begin{align}
  \label{eq:6}
  Q_{13}&=3,-3, & \mathcal{L}_{23}=&L_{23}+l_{23}=2,-2\,,
\end{align}
and those given in Table \ref{tab:1}.  According to Ref. \cite{Dreiner:2003hw} the
negative values in Eq. (\ref{eq:6}) do not yield correct quark mass
matrices and therefore we will not consider them.
  \begin{table}[t]
    \centering
    \begin{tabular}{cccccc}\hline
      $Q_{23}$&$d_{13}$&$d_{23}$&$u_{13}$&$u_{23}$&$\mathcal{L}_{13}$\\\hline
      $2$&$4-Q_{13}$&$0$&$8-Q_{13}$&$2$&$7-\mathcal{L}_{23}$\\\hline    
    \end{tabular}
    \caption{Standard model fields $H$--charges differences. Here $\mathcal{L}_{i3}=L_{i3}+l_{i3}$}
    \label{tab:1}
  \end{table}

\item Two additional phenomenological constraints corresponding to the
  absolute value of the third generation fermion masses,
  $m_t\simeq\langle H_u \rangle$ and $m_b\simeq m_\tau$.
\item Three theoretical restrictions resulting from anomaly
  cancellation~through the Green-Schwarz mechanism, namely, two
  Green-Schwarz mixed linear anomaly cancellation conditions, with
  canonical gauge unification $g_3^2=g_2^2=(5/3)g_1^2$, and the
  mixed quadratic anomaly vanishing on its own \cite{Ibanez:1992fy}.
\end{itemize}
Given the above set of conditions 13, out of 17 $H$--charges are
constrained and can be expressed in terms of the remaining 4 that we
choose to be the lepton number violating bilinear $H$--charges $n_i$ and
$x$. When doing so, in addition to the constraint $n_0=-1$, the
expressions for the standard model field $H$--charges shown in Table
\ref{tab:ic} result \cite{Mira} \footnote{We have fixed a global sign
  misprint on $Q_3$ in Ref. \cite{Mira}}.  Note that $n_0=-1$
implies, according to Eq.  (\ref{eq:2}), $\mu_0\sim m_{3/2}\,\theta$
thus yielding a solution to the $\mu$ problem \cite{Nir:1995bu}.
\begin{table}[t]
  \centering
  \begin{tabular}{rl}\hline\\
    $Q_3=$&$\displaystyle -\frac{-3 x (x+10)+(x+4) n_1+(x+7) n_2+(x+9)
        n_3-67}{15 (x+7)}$\\[0.5cm]
    $L_3=$&$\displaystyle \frac{2 (x+1) (3 x+22)-(2 x+23) n_1-2 (x+7)
      n_2+(13 x+97) n_3}{15 (x+7)}$\\[0.5cm]
    $L_2=$&$L_3+n_2-n_3$\\
    $L_1=$&$L_3+n_1-n_3$\\
    $H_u=$&$n_3-L_3$\\
    $H_d=$&$-1-H_u$\\
    $u_3=$&$-Q_3-H_u$\\
    $d_3=$&$-Q_3-H_d+x$\\
    $l_3=$&$-L_3-H_d+x$\\
    $Q_1=$&$3+Q_3$\\
    $Q_2=$&$2+Q_3$\\
    $u_1=$&$5+u_3$\\
    $u_2=$&$2+u_3$\\
    $d_1=$&$1+d_3$\\
    $d_2=$&  $d_3$\\
    $l_1=$&  $5-n_1+n_3+l_3$\\
    $l_2=$&  $2-n_2+n_3+l_3$\\\\\hline
  \end{tabular}
  \caption{Standard model fields $H$--charges in terms of 
    the bilinear $H$--charges $n_i$ and $x$}
  \label{tab:ic}
\end{table}

As can be seen from Table \ref{tab:ic} the $H$--charges $n_i$ and $x$ act
as free parameters and their possible values should be fixed by
additional experimental constraints.  Mostly motivated by the fact
that $R$--parity breaking models provide a consistent framework for
neutrino masses and mixings \cite{Dreiner:2003hw,Choi:1998wc,Mira,Dreiner:2006xw}, so far in models based on a
single $U(1)_H$ Abelian symmetry the $n_i$ charges have been fixed by
using neutrino experimental data. Here, as already mentioned, we argue
that another approach can be followed by requiring a long--lived, but
not absolutely stable, neutralino. Astrophysical and cosmological
observations require the neutralino decay lifetime to be much more
larger than the age of the Universe
\cite{Barbieri:1988mw,Berezinsky:1996pb,Baltz:1997ar,Arvanitaki:2008hq} which is completely
consistent with the value required to explain the recent reported
anomalies in cosmic-ray electron/positron fluxes ($\tau_\chi\gtrsim
10^{26}$ sec) through decaying DM~\cite{Nardi:2008ix}. Certainly true,
such a long--lived neutralino will be possible only if the couplings
governing its decays are sufficiently small.

If neutralino decays are induced by bilinear $R$--parity violating
couplings, the constraint on $\tau_\chi$ will enforce the ratio
$\mu_i/\mu_0$ to be below $\sim 10^{-23}$ \cite{Huber:2005iz}. Whether
such a bound can be satisfied will depend upon the values of the $n_i$
charges that once fixed will determine the fermion $H$--charges and a
viable model will result if the condition $|f_i|< 10$ can be satisfied
as we already discussed.  Consider the case $n_i<0$: The constraint
$\mu_i/\mu_0$ implies, according to Eq.~(\ref{eq:2}),
$\mu_i/\mu_0\sim\theta^{|n_i|-1}\sim\theta^{34}$ and thus $n_i=-35$.
With these values and from the setup of equations in Table \ref{tab:ic} we
have found that in this case the largest fermion $H$--charge is a
fraction close to 21 in clear disagreement with the condition $|f_i|<
10$. In the case $n_i\ge 0$ the suppression has to be much more stronger
implying larger values for $n_i$ and correspondingly larger values for
$|f_i|$.  Consequently, consistency with astrophysical and cosmological
data excludes the possibility of neutralino decays induced by bilinear
$R$--parity breaking couplings and therefore fix the $n_i$ charges to be
fractional.

Once the $n_i$ charges are chosen to be fractional we are left with a
purely trilinear $R$--parity violating framework in which the order of
magnitude of the couplings is constrained by $n_i$. Unavoidably,
fractional $n_i$ charges imply vanishing $\lambda_{ijj}$ and
$\lambda'$ [see Eqs. (\ref{eq:4}) and (\ref{eq:3}) in the Appendix].
The other couplings [$\lambda_{ijk}$ ($i\ne j\ne k$) and $\lambda''$] can
vanish or not depending on the values of the $n_i$ charges which are
arbitrary as long as they satisfy the constraint $|f_i|<10$. This
freedom allows one to define a set of models which we now discuss in turn:
\begin{enumerate}[($i$)]
\item Models in which the fractional $n_i$ charges are such that all
  the trilinear $R$--parity violating couplings are forbidden as well, and
  the MSSM is obtained \cite{Dreiner:2003yr,Dreiner:2007vp}.  This can be achieved for example by
  fixing $n_1=-3/2$, $n_2=-5/2$ and $n_3=-5/2$ \cite{Dreiner:2003yr}.
\item Models with only a single nonvanishing $\lambda_{ijk}$ and
  vanishing $\lambda''$ couplings. Let us discuss this in more detail.
  An integer $n_{\lambda_{ijk}}$ will imply a nonvanishing
  $\lambda_{ijk}$ coupling. The $H$--charges of the remaining couplings
  ($n_{\lambda''}$, $n_{\lambda_{jki}}$, $n_{\lambda_{ikj}}$) can be
  determined from Eq. (\ref{eq:9}), which can be rewritten as
  \begin{equation}
    \label{eq:7}
    n_i=n_{\lambda_{ijk}}-x-\mathcal{L}_{k3}-1-n_{jk} \qquad (i\ne j\ne k)\,.
  \end{equation}
  From this expression the sum of $n_i$ charges ($N=n_i + n_j + n_k$),
  that according to Eq. (\ref{eq:8}) determine the $n_{\lambda''}$, become
  \begin{equation}
    \label{eq:N-rewritten}
    N=n_{\lambda_{ijk}} - x - \mathcal{L}_{k3} - 1 + 2n_k\,,
  \end{equation}
  and from Eqs. (\ref{eq:7}) and (\ref{eq:9}) the other trilinear
  charges ($n_{\lambda_{jki}}$ and $n_{\lambda_{ikj}}$) can be
  expressed as
  \begin{align}
    \label{eq:TRpV-lam-charges-1}
    n_{\lambda_{jki}}& = -n_{\lambda_{ijk}}+\mathcal{L}_{i3}+\mathcal{L}_{k3}+2x+2+2n_j\,,\\
    \label{eq:TRpV-lam-charges-2}
    n_{\lambda_{ikj}}& =
    n_{\lambda_{ijk}}-\mathcal{L}_{k3}+\mathcal{L}_{j3}-2n_{jk}\,.
  \end{align}
  Thus, from the set up of Eqs. (\ref{eq:N-rewritten}),
  (\ref{eq:TRpV-lam-charges-1}) and (\ref{eq:TRpV-lam-charges-2}) it
  can be seen that as long as $n_j$, $n_k$ and $n_{jk}$ are not
  half-integers an integer $n_{\lambda_{ijk}}$ charge enforces $N$,
  $n_{\lambda_{jki}}$, and $n_{\lambda_{ikj}}$ to be fractional which
  implies that all the $\lambda''$ as well as any other $\lambda$
  coupling different from $\lambda_{ijk}$ vanishes.  Consequently, in
  models in which there is an integer trilinear charge $n_\lambda$ and
  the charges $n_j$, $n_k$ and $n_{jk}$ are not half-integers only a
  single $\lambda$ coupling is allowed.
\item Models in which a single $\lambda_{ijk}$ and all the $\lambda''$
  are nonvanishing. As in the previous case $n_{\lambda_{ijk}}$ must be
  an integer and in addition the corresponding $n_k$ must be a
  half-integer as to guarantee an integer $N$ (see Eq.
  \eqref{eq:N-rewritten}). Moreover, $x$ and the resulting $N$ should
  {\it conspire} to yield a set of integer $n_{\lambda''}$ charges
  (see Eq. \eqref{eq:8}).
\item Models with nonvanishing $\lambda_{ijk}$ and $\lambda_{jki}$.
  These models result once the $n_{\lambda_{ijk}}$ is an integer and
  $n_j$ a half-integer. As can be seen from Eq.
  \eqref{eq:TRpV-lam-charges-1} in this case $n_{\lambda_{jki}}$
  turn out to be an integer allowing the $\lambda_{jki}$ as required.
  Nonvanishing $\lambda_{ijk}$ and $\lambda_{ikj}$ are also possible
  but never the three couplings simultaneously.
\item Along similar lines as those followed in ($ii$), it can be shown
  that models including only $B$-violating couplings can be properly defined
  once any $n_{\lambda''}$ becomes an integer and the bilinear charges $n_j$,
  $n_k$ and $n_j + n_k$ turn out to be not half-integers.
\end{enumerate}
All the models described above have phenomenological implications. For
instance, the family of models discussed in  ($v$) might lead to a
neutralino decaying hadronically whereas those discussed in
($ii$) and ($iv$) share the property of a leptonically decaying neutralino.
These phenomenological aspects can have interesting consequences for
collider experiments but here we will not deal with them. Instead, in
light of the recent data on cosmic-ray electron/positron fluxes, we
will analyze the phenomenology of a neutralino decaying DM in the
minimal trilinear $R$--parity violating models outlined in ($ii$).

A few additional comments regarding these models are necessary.
Bilinear $R$--parity violating couplings are always induced through RGE
running of the trilinear breaking parameters \cite{Nardi:1996iy}:
\begin{align}
  \label{eq:667}
  \mu_i = \frac{\mu_0}{16\pi^2}
  \left[\lambda_{ijk}
    \left(\boldsymbol{h}_e^*\right)_{jk}
    + 3\lambda'_{ijk}\left(\boldsymbol{h}_d^*\right)_{jk}
  \right]\ln\left(\frac{M_X}{M_S}\right)\,,
\end{align}
where $M_X$ is the scale that defines the purely trilinear model and
$M_S$ is the scale of the supersymmetric scalars. At first sight these
parameters could render the minimal trilinear $R$--parity violating
models valid only when the corresponding trilinear couplings are
sufficiently small so to guarantee that the astrophysical and
cosmological bounds on the neutralino lifetime are satisfied.
However, since the $\lambda'$ and $\lambda_{ijj}$ couplings are always
vanishing and the contributions of the allowed $\lambda_{ijk}$ ($i\ne
j\ne k$) require nondiagonal $h_{jk}^e$ which are forbidden by
$U(1)_H$, in these kind of models no bilinear parameters can be
induced at all.

\section{Decaying neutralino dark matter }
\label{sec:decaying-neutralino}
\begin{table}[t]
  \centering
  \begin{tabular}{lccccc}\hline
    & $x$& $n_1$   &$n_2$      &$n_3$     & $|f_i|$\\\hline
    $\lambda_{231}$& $1$& $7/3$ & $-19/3$  & $-25/3$&  $<7$ \\\hline 
    $\lambda_{123}$& $1$& $-10/3$ & $-19/3$  & $7/3$&  $<6$ \\\hline 
    $\lambda_{132}$& $1$& $-5/3$ & $17/3$  & $-20/3$&  $<7$ \\\hline 
  \end{tabular}
  \caption{Set of bilinear $H$--charges consistent with the trilinear 
    $H$--charge choice $n_\lambda=-10$.}
  \label{tab:mneg}
\end{table}
In this section we will study neutralino decays in the context of the
minimal $R$--parity violating models that were defined in the previous
section.  The lifetime of a mainly gaugino neutralino decaying through
a trilinear $R$--parity breaking coupling $\lambda$ is approximately
given by \cite{Baltz:1997ar}
\begin{align}
\label{tau_gau}
\tau_{\chi}\sim  \left(\frac{M_S}{2\times10^4\text{GeV}}\right)^4
\left(\frac{10^{-23}}{\lambda} \right)^2
\left(\frac{2\times10^3\text{GeV}}{m_\chi}\right)^5\;10^{26}\;\text{sec}\,.
\end{align}
According to this expression the viability of a neutralino decaying DM
will depend, for a few TeV neutralino mass, on the slepton mass
spectrum and the size of the corresponding $\lambda$ coupling that
will be determined by the choices $n_\lambda<0$ or $n_\lambda\ge 0$.
These choices are to some extent not arbitrary as they must satisfy
the condition $|f_i|<10$: Given a value for $n_\lambda$, the $n_j$ and
$n_k$ charges can be fixed through Eq.~(\ref{eq:7}) and for a
particular $x$ the different $|f_i|$ charges can be calculated. Tables
\ref{tab:mneg} and \ref{tab:mpos} show some examples.

\begin{figure}[t] 
 \includegraphics{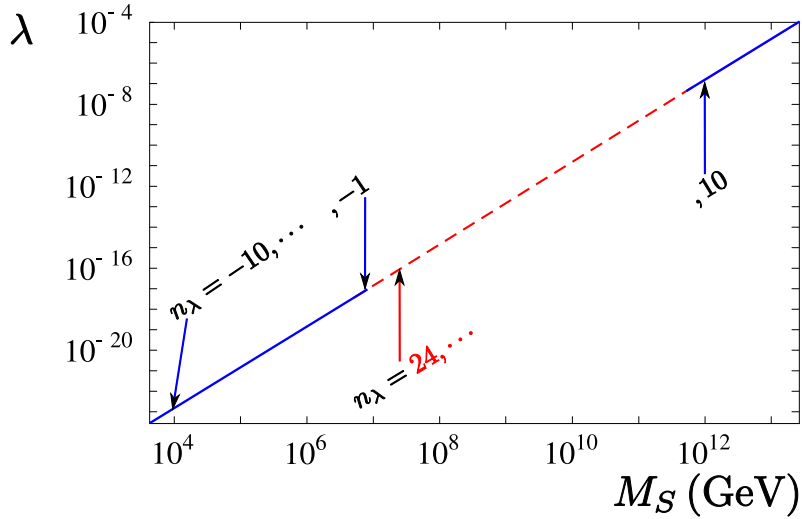}
 \caption{$\lambda$ coupling as a function of the slepton mass for a
   neutralino lifetime of $10^{26}$ sec The solid lines (blue)
   correspond to values of $\lambda$ and $M_S$ well given by a minimal
   trilinear $R$--parity breaking model whereas those in the range of the
   dashed line (red) are not consistent with the limits on $H$--charges
   (see text for details).}
\label{fig:lambda}
\end{figure}

In the case $n_\lambda<0$, due to the strong suppression induced by
the factor $m_{3/2}/M_P$, a coupling $\lambda$ as small as $10^{-23}$
is possible if $n_\lambda=-10$ and accordingly even with a not so
heavy slepton the constraint $\tau_\chi=10^{26}$ sec can be
satisfied.  In the case $n_\lambda\ge 0$ such a small $R$--parity breaking
coupling will require a value for $n_\lambda$ irreconcilable with the
limit $|f_i|<10$.  Thus, in this case $\lambda$ will be larger and the
correct neutralino lifetime will result, if possible, only from an
additional suppression given by a superheavy slepton as those featured
by split supersymmetry \cite{split1}.  Figure \ref{fig:lambda} shows the values of
$\lambda$ (arising from different $n_\lambda$ choices) and $M_S$
consistent with $\tau_\chi= 10^{26}$ sec In the solid line (lower
left corner) $n_\lambda=-10,\dots ,-1$.  In the dashed line
$n_\lambda>10$ and the resulting $|f_i|$ charges are inconsistent with
the requirement $|f_i|<10$ thus ruling out the possibility of a
decaying neutralino dark matter in the range $M_S=10^7-10^{12}$ GeV.
Finally, in the solid line (upper right corner) $n_\lambda=6,\dots,
10$. The values below 6 will require a slepton mass above $10^{13}$~GeV that leads to a gluino lifetime exceeding the age of the Universe
\cite{Bernal:2007uv} and therefore are excluded.
\begin{table}
  \centering
  \begin{tabular}{lccccc}\hline
      & $x$&  $n_1$   &$n_2$      &$n_3$     & $|f_i|$\\\hline
$\lambda_{231}$& $1$& $-2/3$ & $-1/3$  & $-1/3$&  $<6$ \\\hline 
$\lambda_{123}$& $1$& $5/3$ & $5/3$  & $-5/3$&  $<5$ \\\hline 
$\lambda_{132}$& $1$& $4/3$ & $-1/3$  & $4/3$&  $<5$ \\\hline 
  \end{tabular}
  \caption{Set of bilinear $H$--charges consistent with the trilinear 
    $H$--charge choice $n_\lambda=7$.}
  \label{tab:mpos}
\end{table}

Some words are in order concerning the minimal trilinear $R$--parity
violating models with superheavy sleptons.  In split supersymmetry the
scalar masses, apart from the Higgs boson, are well above the
electroweak scale, $M_S\lesssim 10^{13}$ GeV. Assuming $M_S=10^{13}$
GeV it can be seen from Fig.~\ref{fig:lambda} that the correct
neutralino lifetime requires couplings of order $10^{-5}$. According
to Eq. (\ref{eq:667}) couplings of this size will lead to
$\mu_i/\mu_0\sim 10^{-12}$ in sharp disagreement with the bounds from
astrophysical and cosmological data \cite{Huber:2005iz}. However, in contrast to
previous analysis of neutralino decaying DM within split supersymmetry
\cite{Chen:2009mj,Gupta:2004dz}, in this context the RGE running of
the trilinear $R$--parity breaking parameters do not induce any $R$--parity
breaking bilinear coupling.
\subsection{PAMELA, Fermi and ATIC anomalies}
\label{sec:pamela-fermi-atic}
In this section we will show that the PAMELA, ATIC and Fermi~LAT data
can be well accounted for by a decaying neutralino DM in the context
of minimal trilinear $R$--parity violating models. In order to fit the
electron-positron fluxes we fix the trilinear $R$--parity breaking
coupling and the neutralino mass and lifetime according to
$\lambda=3.2\times 10^{-23}$, $m_\chi=2038$ GeV and
$\tau_\chi=1.3\times 10^{26}$ sec Note that such a coupling can arise
from $n_\lambda=-10$. We generate a supersymmetric spectrum with {\tt
  SuSpect} \cite{Djouadi:2002ze} by choosing the benchmark point $F$
defined in Ref.~\cite{Yin}. In the resulting spectrum the neutralino
becomes mainly wino and the scalar masses have a size of $M_S\sim
10^{4}$~GeV. For cosmic rays propagation, we followed
Ref.~\cite{Cirelli:2008id} whereas for DM we used the spherically
symmetric Navarro, Frenk, and White \cite{Navarro:1995iw} profile and
the propagation model MED introduced in Ref.  \cite{Delahaye:2007fr}.
The electron and positron energy spectra were generated using {\tt
  PYTHIA} \cite{Sjostrand:2006za}.
\begin{figure} 
\begin{center}
 \includegraphics[width=6.7cm,height=5.4cm]{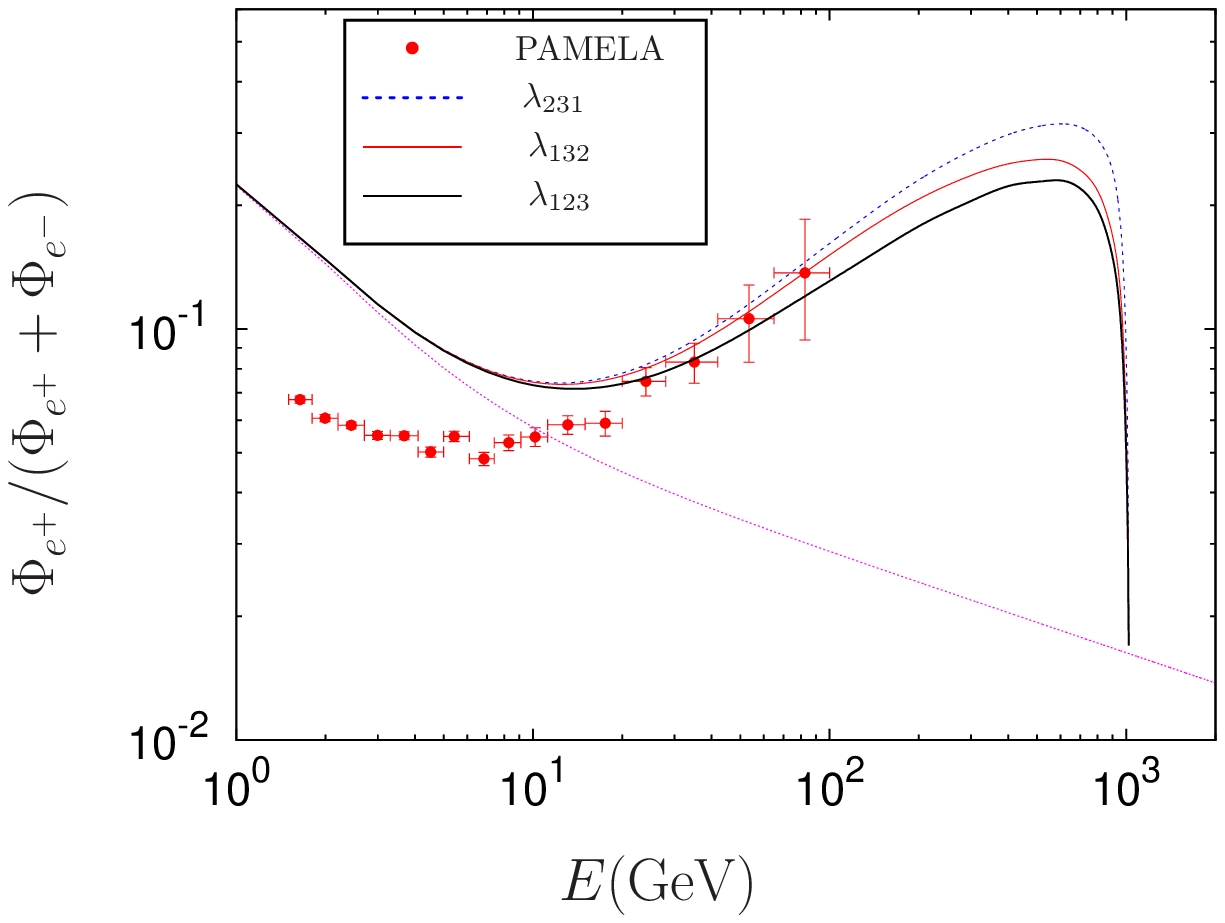}
 \includegraphics[width=6.7cm,height=5.4cm]{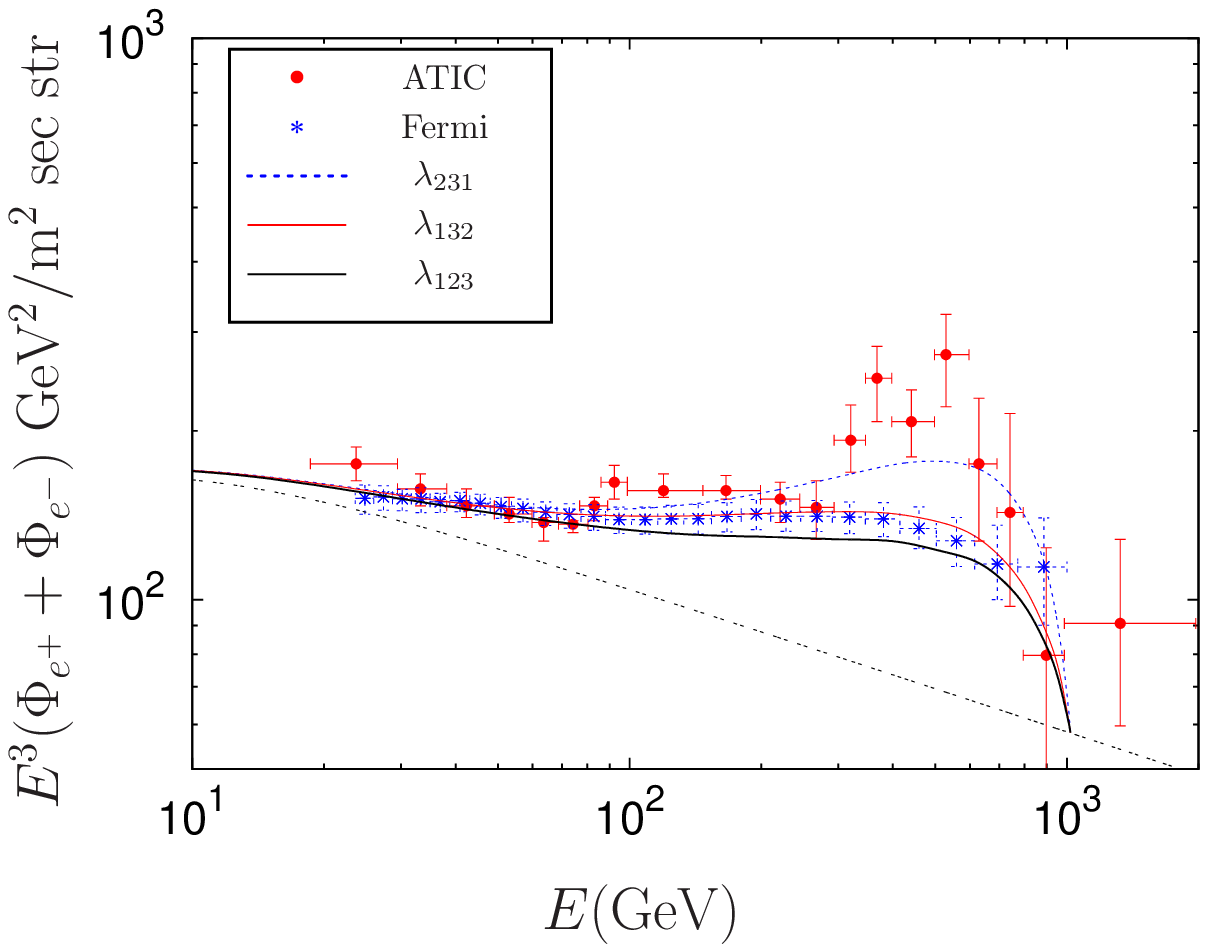}
 \caption{Ratio of positron to electron-plus-positron (left panel) and
   total electron-plus-positron (right panel) fluxes arising from a
   long--lived neutralino decaying through trilinear $R$--parity breaking
   couplings (see text for details).}
\label{fig:pamela}
\end{center}
\end{figure}

Figure \ref{fig:pamela} shows the ratio of positron to
electron-plus-positron and the total electron-plus-positron fluxes
originated from neutralino decays induce by the trilinear $R$--parity
violating couplings $\lambda_{231}$, $\lambda_{132}$ and
$\lambda_{123}$. Decays induced by $\lambda_{231}$ always involve hard
electrons and positrons and therefore are well suited to explain ATIC
data as can be seen in Fig.~\ref{fig:pamela}. In contrast, the decays
through the couplings $\lambda_{132}$ and $\lambda_{123}$ involve
either final state muons or taus and thus electron and positrons with
lower energies as those required to explain Fermi~LAT and PAMELA
measurements. Once the neutralino mass is fixed its lifetime will
depend only on the ratio $M_S^2/\lambda$ [see Eq. (\ref{tau_gau})] and
of course will not change as long as this ratio remains constant. Moreover, the effect of  $M_S$ on dark matter relic density is completely negligible for $M_S> 10^{4}\,$GeV \cite{Bernal:2007uv,Pierce:2004mk}.
Accordingly, the results in Fig. \ref{fig:pamela} also hold in the
case of superheavy sleptons and {\it large} couplings.
\section{Conclusions}
\label{sec:summary}
We have studied decaying neutralinos as DM candidates in the context
of supersymmetric models extended with an anomalous $U(1)_H$ flavor
symmetry.  We have shown that theoretical motivated limits on the
standard model field $H$--charges in addition to astrophysical and
cosmological constraints on neutralino decaying DM forbid the decays
induced by bilinear $R$--parity breaking couplings and allow one to define a
set of purely trilinear $R$--parity violating models in which the
neutralino can be stable on cosmological scales. Among all these
scenarios we have found a class of models (minimal $R$--parity violating
models) in which a single $R$--parity and lepton number breaking coupling
$\lambda_{ijk}$ ($i\ne j\ne k$) give rise to leptonic neutralino decays.
In these schemes, for a few TeV neutralino mass, a decaying lifetime
of $\tau_\chi\sim 10^{26}$ sec  can be readily achieved for a variety
of $H$--charge assignments and slepton masses ranging from few TeV up to
the typical scales of split supersymmetry.
 
Moreover, we have shown that these minimal $R$--parity violating models
(depending on the trilinear $R$--parity breaking parameter defining the
model itself) provide an explanation to the observed anomalies in the
electron-positron fluxes reported by PAMELA, ATIC, and Fermi~LAT.
\section*{Acknowledgments}
Work partially supported by CODI-UdeA under Contract No. IN564-CE

\appendix{}
\section{$H$--charges of the different couplings}
\label{sec:solution}
In this Appendix we give expressions for the standard model Yukawa
couplings as well as for the lepton and baryon number couplings
appearing in the superpotential. The quark Yukawa couplings can be
written as
\begin{align}
  h^u_{ij}&\sim\theta^{{Q_{i3}}+{u_{j3}}} &   h^d_{ij}&\sim\theta^{{Q_{i3}}+{d_{j3}}+x}\,,
\end{align}
while the charged lepton Yukawa couplings we have 
\begin{align}
  n_{h^l_{ij}}&={{L_{i3}}+{l_{j3}}+x}\Rightarrow h^l_{ij}\sim 
  \begin{cases}
   \theta^{{L_{i3}}+{l_{j3}}+x} & {{L_{i3}}+{l_{j3}}+x}\geq0\\
   \frac{m_{3/2}}{M_p}\theta^{|{L_{i3}}+{l_{j3}}+x|} & {{L_{i3}}+{l_{j3}}+x}<0\\
   0& {{L_{i3}}+{l_{j3}}+x}\quad \text{fractional}\,.\\
  \end{cases}
\end{align}
Here $n_{h_{ij}^l}$ denotes the $H$--charge of the gauge invariant term
with coupling $h_{ij}$.  From the expressions in tab. \ref{tab:ic} these $H$--charges
can be rewritten in terms of $n_i$ and $x$ as follows:
\begin{align}
n_{h_{ij}^l}=&  L_{i3}+l_{j3}+x\nonumber\\
=&n_{ij} + \mathcal{L}_{j3}+x\,,
\end{align}
where $n_{ij}=n_i - n_j$. For the lepton number and $R$--parity breaking
couplings one can proceed along similar lines, that is to say from the
$H$--charges of the standard model fields involved in each case, and
according to Table~\ref{tab:ic}, the following relations can be
derived:
\begin{align}
\label{eq:9}
 n_{\lambda_{ijk}}=&L_i+L_j+l_k\nonumber\\
=&n_i + n_{jk} + \mathcal{L}_{k3} + 1 + x\,,
\end{align}
and
\begin{align}
  n_{\lambda_{ijk}'}=&L_i+Q_j+d_k\nonumber\\
  =&n_i - n_0 + n_{h^d_{ij}}\,.
\end{align}
Explicitly, the quark Yukawa couplings matrices can be written---up
to order 1 factors---as
\begin{align}
  h^u\sim&\begin{pmatrix}
    \theta^8 & \theta^5 &\theta^3\\ 
    \theta^7 & \theta^4 &\theta^2\\ 
    \theta^5 & \theta^2 &1\\ 
  \end{pmatrix} & 
  h^d\sim&\theta^x\begin{pmatrix}
    \theta^4 & \theta^3 &\theta^3\\ 
    \theta^3 & \theta^2 &\theta^2\\ 
    \theta & 1 &1\\ 
  \end{pmatrix}\,, 
\end{align}
and the charged lepton $H$--charges as
\begin{align}
  n_{h^l_{ij}}=
  \begin{bmatrix}
    x+5 & x+n_1-n_2+2 & x+n_1-n_3 \\
    x-n_1+n_2+5 & x+2 & x+n_2-n_3 \\
    x-n_1+n_3+5 & x-n_2+n_3+2 & x
  \end{bmatrix}\,.
\end{align}
The trilinear lepton number violating couplings $H$--charges are
given by
\begin{align}
\label{eq:4}
  &\begin{bmatrix}
    n_{\lambda_{211}} &  n_{\lambda_{212}} & n_{\lambda_{213}} \\
    n_{\lambda_{311}} &  n_{\lambda_{312}} & n_{\lambda_{313}} \\
    n_{\lambda_{231}} &  n_{\lambda_{232}} & n_{\lambda_{233}} \\
  \end{bmatrix}=\nonumber\\
  &\begin{bmatrix}
    {x+n_2+6} & {x+n_1+3} & {x+n_1+n_2-n_3+1} \\
 {x+n_3+6} & {x+n_1-n_2+n_3+3} & {x+n_1+1} \\
 {x-n_1+n_2+n_3+6} & {x+n_3+3} & {x+n_2+1}
  \end{bmatrix}
\end{align}
and
\begin{align}
  \label{eq:3}
  n_{\lambda_{ijk}'}=
  \begin{bmatrix}
    x+n_i+5 & x+n_i+4 & x+n_i+4 \\
    x+n_i+4 & x+n_i+3 & x+n_i+3 \\
    x+n_i+2 & x+n_i+1 & x+n_i+1
  \end{bmatrix}\,,
\end{align}
Finally for the baryon number breaking couplings we found
\begin{align}
\label{eq:8}
  &\begin{bmatrix}
    n_{\lambda_{121}''} &  n_{\lambda_{221}''} & n_{\lambda_{321}''} \\
    n_{\lambda_{131}''} &  n_{\lambda_{231}''} & n_{\lambda_{331}''} \\
    n_{\lambda_{123}''} &  n_{\lambda_{223}''} & n_{\lambda_{323}''} \\
  \end{bmatrix}=\nonumber\\
  &\begin{bmatrix}
    {\frac{1}{3} \left(3 x+N+17\right)} &
   {\frac{1}{3} \left(3 x+N+8\right)} &
   {\frac{1}{3} \left(3 x+N+2\right)} \\
 {\frac{1}{3} \left(3 x+N+17\right)} &
   {\frac{1}{3} \left(3 x+N+8\right)} &
   {\frac{1}{3} \left(3 x+N+2\right)} \\
 {\frac{1}{3} \left(3 x+N+14\right)} &
   {\frac{1}{3} \left(3 x+N+5\right)} &
   {\frac{1}{3} \left(3 x+N-1\right)}
  \end{bmatrix}\,,
\end{align}
where we have defined $N=n_1+n_2+n_3$.

\end{document}